\pdfoutput=1
\documentclass[a4paper,10pt,amsmath,amssymb,notitlepage,twoside,superscriptaddress,nofootinbib,eqsecnum,aps,prd]{revtex4-1}
\usepackage[latin1]{inputenc}
\usepackage[T1]{fontenc}
\usepackage[usenames]{color}
\usepackage[final]{showkeys} 
\usepackage{enumerate}
\usepackage[pdftex]{hyperref}
\hypersetup{colorlinks=true,linkcolor=black,citecolor=black,filecolor=black,urlcolor=black,
            pdfauthor={M. Cadoni, E. Franzin and M. Serra},
            pdftitle={Horizons, conical singularities and holographic entanglement entropy}}
\usepackage[capitalise]{cleveref}
\usepackage{tikz}
\usetikzlibrary{shapes}

\crefformat{plural}{#2Eqs.~(#1)#3} 
\crefname{section}{Sect.}{Sects.}

\makeatletter\g@addto@macro\bfseries{\boldmath}\makeatother%
\def\beq{\begin{equation}}
\def\eeq{\end{equation}}

\renewcommand{\leq}{\leqslant}

\renewcommand{\le}{\leqslant}

\def\0{\nonumber}

\def\lb{\label}

\def\a{\alpha}
\def\b{\beta}
\def\lt{l_t}
\def\ls{l_s}

\def\be#1\ee{\begin{align}#1\end{align}}
\def\bsube[#1]#2\esube{\begin{subequations}\label[plural]{#1}\begin{align}#2\end{align}\end{subequations}}

\begin{document}

\title{\texorpdfstring{How  is the Presence of  Horizons and Localised Matter Encoded in the Entanglement Entropy?}{How  is the Presence of  Horizons and Matter Encoded in the Entanglemnt Entropy?}}
\author{Mariano Cadoni}\email{mariano.cadoni@ca.infn.it}
\affiliation{Dipartimento di Fisica, Universit\`a di Cagliari,\\Cittadella Universitaria, 09042 Monserrato, Italy}
\affiliation{INFN, Sezione di Cagliari}
\author{Parul Jain }\email{parul.jain@ca.infn.it}
\affiliation{Dipartimento di Fisica, Universit\`a di Cagliari,\\Cittadella Universitaria, 09042 Monserrato, Italy}
\affiliation{INFN, Sezione di Cagliari}
\date{\today}

\begin{abstract}
Motivated by the new theoretical paradigm  that views spacetime geometry as emerging from the entanglement of a pre-geometric theory, we 
investigate the issue of the signature of the presence of horizons and localized matter on the entanglement entropy (EE) $S_E$ for the case of three-dimensional AdS (AdS$_3$) gravity. We use the  holographically dual two-dimensional  CFT on the torus and the related modular symmetry in order to treat bulk black holes and conical singularities (sourced by pointlike masses not shielded by horizons)   on the same footing.
In the regime where  boundary tori can be approximated by cylinders we are able to give universal expressions for the EE  of  black holes and conical singularities. We argue that the presence of horizons/localized matter in the bulk is encoded in the EE in terms of 
$(i)$ enhancement/reduction of the entanglement of the AdS$_3$ vacuum, 
$(ii)$  scaling as area/volume of the leading term of the perturbative 
expansion of $S_E$, $(iii)$ exponential/periodic behaviour of $S_E$, $(iv)$ presence of  unaccessible regions   in the noncompact/compact dimension of the boundary cylinder.
In particular, we show that  the reduction effect  of matter on the entanglement of the vacuum found by Verlinde  for the de Sitter vacuum extends to the AdS$_3$ vacuum. 
\end{abstract}

\maketitle
\tableofcontents
\section{Introduction}
\lb{sec1}
One of the most  fruitful, recently proposed,  concepts  in the context of fundamental gravitational physics is that spacetime geometry and gravity emerge from quantum entanglement of a microscopic,  pre-geometric theory \cite{Jacobson:1995ab,Padmanabhan:2009vy,Verlinde:2010hp,VanRaamsdonk:2010pw,Casini:2011kv,Faulkner:2013ica,Swingle:2014uza,Jacobson:2015hqa,Verlinde:2016toy}. 
It has been shown that the connectivity of the spacetime can be explained in terms of quantum entanglement \cite{VanRaamsdonk:2010pw}. It has been proved that entanglement entropy (EE)  satisfies the same  thermodynamical relations of the Bekenstein-Hawking  entropy \cite{Lewkowycz:2013nqa,Casini:2011kv}. The linearized  Einstein equations have been 
derived from  quite general principles involving quantum entanglement \cite{Faulkner:2013ica,Swingle:2014uza,Jacobson:2015hqa}. It has been also proposed that the accelerated expansion of the universe can be explained  by  a volume law contribution to the entanglement entropy producing a slow thermalization of the emergent spacetime 
\cite{Verlinde:2016toy}. In this framework the phenomenology commonly attributed to dark matter  can be explained as a "dark force" originated by competition between 
area and volume contribution to EE \cite{Verlinde:2016toy}.

This novel paradigm  has been triggered by  the advances in the understanding of the microscopic  origin of the Bekenstein-Hawking entropy \cite{Strominger:1996sh,Cadoni:1998sg}, by the development of the AdS/CFT correspondence \cite{Maldacena:1997re} and the related holographic interpretation of entanglement entropy \cite{Ryu:2006bv,Ryu:2006ef}, by earlier work on the emergence of gravity \cite{Jacobson:1995ab} and, last but not the least,  by the recent discussions about the black hole quantum information puzzle involving firewalls and complementarity \cite{Almheiri:2012rt,Maldacena:2013xja,Susskind:2014moa}.

A weak  point of this new emergent gravity  scenario, which in our opinion has not been adequately  stressed,  is that it requires for its 
formulation  a background pre-existing geometric structure, like e.g. local  Killing \cite{Jacobson:1995ab,Faulkner:2013ica,Swingle:2014uza,Jacobson:2015hqa}   or cosmological \cite{Verlinde:2016toy} horizons. 
This is conceptually non consistent because one would like to  have a background-independent formulation, in which gravity and spacetime emerges 
completely from a pre-geometric microscopic structure.  This seems to support the point  of view, according to which entanglement is not enough  to explain all the  quantum information puzzles  and that some additional crucial feature is needed \cite{Susskind:2014moa}.

A promising way to try to shed light on this issue is to look at the way the information 
about the presence of background gravitational structures (horizons, vacuum spacetimes, spacetimes  with  singularities)
is encoded in the  EE. This is the main purpose of this paper.
We investigate this issue in the context of AdS gravity in three spacetime dimensions (3D),
in  which pointlike matter generates conical singularities. Moreover in 3D,
the AdS/CFT correspondence, the  Ryu-Takayanagi (RT) formalism \cite{Ryu:2006bv,Ryu:2006ef,Azeyanagi:2007bj}, and its  extensions like  complexified geodesic \cite{Balasubramanian:2012tu} and  differential entropy \cite{Balasubramanian:2014sra} allows us to compute in closed form the EE of the dual two-dimesional (2D) Euclidean  CFT living on the torus. 

The modular symmetry of the torus will  be used to exchange regions  not accessible to measurement, 
to which an EE $S_E$ is associated, between different directions of Euclidean space and   to treat bulk geometries with horizons and conical singularities on the same footing. 
After that, we work in the limit where the boundary tori can be approximated by  cylinders. 
In this regime the modular symmetry will provide us with  a web of dualities mapping the various boundary cylinders 
with unobservable regions. Moreover, in the cylinder  approximation the EE of the Euclidean boundary CFT belongs to two 
universality classes corresponding to  unobservable slices in the  compact or noncompact dimension. This will allow us to assign a well-defined EE to the various gravitational bulk configurations and to discuss the signature corresponding to the presence of horizons or pointlike masses not shielded by an event horizon.

We find  that  the presence of an event horizons in the bulk is related to  an enhancement of 
the entanglement of the AdS$_3$ vacuum, to an area law for  the leading term of the perturbative expansion of $S_E$, 
to the  presence of unobservable regions   in the 
noncompact dimension of the cylinder   and to exponential behaviour of the EE. Conversely,  
localized  matter not shielded by an event horizon, i.e  conical singularities in the bulk,  are related  to 
reduction of the entanglement of the AdS$_3$ vacuum, to a volume law for  the leading term of the perturbative expansion 
of $S_E$, to the  presence of  unobservable regions   in the
compact dimension and to   a periodic behaviour of the EE.  In particular, this  confirms the reduction effect  of matter on the entanglement of the vacuum found in Ref. \cite{Verlinde:2016toy} for de Sitter spacetime.

We also  briefly discuss EE in Minkowski spacetime,  the concept of entwinement \cite{Balasubramanian:2014sra}, complexified geodesic \cite{Balasubramanian:2012tu} and  the concept of entanglement in time.

The structure of this paper is as follows. We start the paper by reviewing some aspects of AdS gravity in 3D and related  bulk configuration (Sect. \ref{sec2}).
We discuss modular transformations in the torus in section \ref{sec3}\hspace{0.05cm}  and their cylinder limit in  Sect. \ref{sec4}.     
Then we move on to section \ref{sec5} where we obtain entanglement entropy  
of the dual  CFT for different locations of the slices in the cylinder approximation. 
In Sect. \ref{sec6} we compute the holographic entanglement entropy of various the 3D gravity bulk solutions. 
In section \ref{sec7}\hspace{0.05cm}, we use gravitational tools to 
cross-check the results of Sect. \ref{sec6}.
In Sect. \ref{sec8} we derive the leading terms for 
the holographic EE  in the large/small radius  expansion associated to  the various gravitational bulk configurations.
In Sect. \ref{sec9} we perform   a Wick rotation to derive   the EE in Minkowski spacetime.
In Sect. \ref{sec10} we briefly discuss  causal aspects of holographic EE.
We conclude the paper by stating our conclusions (Sect. \ref{sec11}).

%
\section{Bulk gravity configurations}
\label{sec2}
In this paper we will restrict our considerations to 3D AdS gravity. 
Classical, pure AdS$_{3}$ gravity is  described by the  action
\beq\lb{e1}
A=\frac{1}{16\pi G_{3}}\int d^{3}x \left(R+ \frac{2}{L^{2}}\right),
\eeq
where $L$ is the de Sitter length
and $G_{3}$ is 3D Newton constant.

The $\mathrm{AdS_{3}/CFT_{2}}$ correspondence \cite{Maldacena:1997re} dictates that in the large ${\cal N}$ limit, i.e in the regime where the central charges  of the $CFT_{2}$ \cite{Brown:1986nw}
\be\lb{cc}
c=\bar c=\frac{3L}{2G_{3}}\gg 1,
\ee
 3D AdS Gravity is holographically dual to 2D CFT defined on the $r\to \infty$ conformal boundary  (torus). 
\subsection{ The spectrum of AdS$_3$ gravity} 
Pure AdS gravity in 3D has no propagating degrees of freedom. Away from the sources the spacetime is always locally equivalent to  AdS$_3$.
Localised matter affects only the global  properties of the spacetime. 
The solutions of AdS$_{3}$ classical gravity, which represent regular geometries, are usually classified in terms of three classes corresponding 
to the orbits (elliptic, hyperbolic, parabolic)
of the $SL(2,R)$ group manifold \cite{Banados:1992wn,Banados:1992gq,Carlip:2005zn}. However, a physical classification of the solutions  can be obtained by starting from the BTZ black hole, i.e an asymptotically AdS solution with an event horizon, positive mass and  finite temperature $T$.
The BTZ black holes give the positive mass excitations. The vacuum of the theory is  the zero mass, $T=0$  solution and it is given by AdS$_{3}$ in Poincar$\mathrm{\acute{e}}$ coordinates.
In this paper we will considers this as the physical vacuum of the theory.  The peculiarity of AdS$_3$ gravity is that in the theory we have an
other vacuum solution, namely  AdS$_{3}$ in  global coordinates. This solution is separated from the AdS$_{3}$ Poincar$\mathrm{\acute{e}}$ vacuum by solutions describing conical singularities. Although these geometries are not regular  we will consider them  as real physical solutions, because differently from  gravity in four dimensions, they are sourced  by a pointlike particle of {\sl positive mass }.

In this paper we will therefore take in considerations  all the solutions of AdS$_{3}$ classical gravity, i.e including also those representing conical singularities in space and time.
As we will explain in the next section this is necessary in view of the  modular symmetry of the  CFT living 
in the boundary torus. For our purposes we will distinguish between four classes of solutions
namely AdS$_{3}$ vacua, 
the  BTZ Black hole and AdS$_{3}$ with conical defect along, respectively, the space and time direction.  
\begin{itemize}
\item
{\bf AdS$_{3}$ vacua}
  
As we have mentioned above, we have two vacua:  AdS$_3$ in global coordinates and AdS$_3$ in the Poincar$\mathrm{\acute{e}}$ patch.

Global AdS$_3$    can be represented by the metric
\begin{equation}\label{6}
ds^2 = -\left(1 + \frac{r^2}{L^2}\right)dt^2 +\left(1 + \frac{r^2}{L^2}\right)^ {-1} dr^2 + \frac{r^2}{L^2}d\phi^2.
\end{equation}
These global coordinates range over $ t \in [-\infty,\infty], \phi \in [0,2\pi L] , r \in [0,\infty)$.

By restricting $ t \in [0,\beta]$ we get   AdS$_{3}$ at finite temperature $T=1/\b$.
This configuration is dual to a  CFT with  the same finite temperature and living on a torus 
$\mathcal{T}$($\beta , 2\pi L$), where $\beta$ and $2\pi L$ represent, respectively, the cycles in the time and space direction.
Also, one can take the Poincar$\mathrm{\acute{e}}$ patch of $\mathrm{AdS_{3}}$ which can be obtained by taking the limit 
$r\rightarrow \infty$ of \eqref{6} and is given by the metric:
\begin{equation}\label{8}
 ds^2 = -\left(\frac{r^2}{L^2}\right)dt^2 +\left(\frac{r^2}{L^2}\right)^ {-1} dr^2 + \frac{r^2}{L^2} d\phi^2
\end{equation}
Here, $\phi$ is no longer periodic, $ \phi \in [-\infty,\infty],\,t \in [-\infty,\infty]$ and this solution is dual to a CFT living on a Plane.

 In  this paper we will consider  AdS$_3$ in the Poincar$\mathrm{\acute{e}}$ patch  as the physical vacuum of the theory. The rational behind our choice is that this solution, differently from global AdS$_3$,  is  continuously  connected both with the BTZ black hole part of the spectrum and with the part of the spectrum describing conical singularities, which are sourced  by a pointlike mass. We can therefore consider both thermal excitation of the vacuum  generated by a black hole with non vanishing mass and insertion of pointlike masses  
producing conical singularities.
\item
{\bf BTZ  black hole}

It can be represented by the metric :
\begin{equation}\label{9}
ds^2 = -\frac{1}{L^2}\left(r^2 - r_{+}^2\right)dt^2 +\left(r^2 - r_{+}^2\right)^ {-1}L^2 dr^2 + \frac{r^2}{L^2}d\phi^2. 
\end{equation}
The BTZ black hole has horizon radius $r_{+}$,  inverse Hawking temperature $\beta_{H}$, 
mass and thermal Bekenstein-Hawking entropy given by 
\beq\label{10}
\beta_{H} = \frac{1}{T_{H}} = \frac{2\pi L^2}{r_{+}},\quad
M = \frac{r_{+}^2}{8G_{3}L^2}, \quad
S_{BH} = \frac{\mathcal{A}}{4G_{3}} = \frac{\pi r_{+}}{2G_{3}}.
\end{equation}
The Poincar$\mathrm{\acute{e}}$ patch of $\mathrm{AdS_{3}}$ can be considered as the ground state of BTZ black hole in which we
have $r_{+} = 0,\, M = 0$ and $T_{H}= 0$. 
The corresponding 3D Euclidean  space has a contractible cycle in the $t$ direction and for generic values of its periodicity $\beta$ there is a 
conical singularity  along the $t$ direction. For $\beta = \beta_{H}$, the conical singularity is removed and we have a
CFT at a finite temperature   living on a torus  $\mathcal{T}$($\beta_{H} , 2\pi L$) with cycles 
$\beta_{H} , 2\pi L$.
\item
{\bf AdS$_{3}$\ with\ conical\ singularity\ in\ space}  

It can be
obtained from Eq. \eqref{6} by rescaling the coordinates as
\begin{equation}\label{11}
\begin{aligned}
t \rightarrow \Gamma t, \quad \phi \rightarrow  \Gamma\phi \quad \mathrm{and} \quad  r \rightarrow r/\Gamma \quad  \mathrm{where}
\quad  \Gamma = \frac{r_{+}}{L}.
\end{aligned}
\end{equation}
With this rescaling we get the following metric:
\begin{equation}\label{12}
 ds^2 = -\left(\Gamma^2 + \frac{r^2}{L^2}\right)dt^2 +\left(\Gamma^2 + \frac{r^2}{L^2}\right)^ {-1} dr^2 
 + \frac{r^2}{L^2}d\phi^2,
\end{equation}
where the coordinates range is $ t \in [0,\infty], \phi \in [0,2\pi L] , r \in [0,\infty]$ and 
with the parameter $0\le{\Gamma}\le 1$ ranging 
from  Poincar$\mathrm{\acute{e}}$\; AdS ($\Gamma=0$) to  global AdS ($\Gamma=1$). For $\Gamma\neq 1,0$ the corresponding 3D euclidean space  has a contractible cycle 
along in the spatial $\phi$-direction. 
If we restrict $\phi$ as 
$0 \leq \phi \leq 2\pi\Gamma L$ where $\Gamma = r_{+}/L$ then we will have a space with a conical singularity owing to the
deficit angle given as\\ $\delta\phi = 2\pi(1- \Gamma)= 2\pi(1- r_{+}/L)= 2\pi(1-2\pi L/\beta_{con})$, where  in analogy
 with the inverse Hawking temperature we  have defined,
\beq\lb{bcon}
\beta_{con} \equiv \frac{2\pi L^2}{r_{+}}=\beta_H.
\eeq
By restricting the coordinates to $t \in [0,\Gamma\beta_{con}=2\pi L], \phi \in [0,2\pi\Gamma L], r \in [0,\infty)$,
we have a spacetime with a conical singularity, whose  $r\to\infty$ conformal boundary is a  torus  $\mathcal{T}$($\Gamma\beta_{con},2\pi\Gamma L$). This $
\mathrm{AdS_{3}}$ space  with conical defect
is dual to a CFT living on  this torus.
The conical singularity we  have whenever $\Gamma\neq 0,1$
represents
the geometric
distortion generated by a pointlike particle of mass
$m=(1-\Gamma)/4G_{3}$ \cite{Deser:1983tn,Deser:1983nh,Brown:1986nw}.
In fact a stress tensor for a pointlike mass $m$ provides a solution of Einstein equations in 3D given by 
Eq. (\ref{12}). It is important to stress that what we are considering as the physical vacuum of the theory (AdS$_3$ in the Poincar$\mathrm{\acute{e}}$ patch)  can be also  obtained from the conical singularity spacetime by setting $\Gamma=0$, which from Eq. (\ref{bcon}) implies $\beta_{con}=\infty$, and by decompactifying the $t$ direction. 
\\

\item
{\bf AdS$_{3}$\ with\ conical\ singularity\ in\ time}

For generic values of the Euclidean time periodicity different from $\beta_H$,   the Euclidean version of the  spacetime   (\ref{9}) describes a conical singularity in the time direction. We can always parametrize this periodicity  in terms of $r_+$ and choose $t\in [0,2\pi r_+= 2\pi \Gamma L]$, where $\Gamma$ is defined as above. 
In this way the conical singularity in time has the same characterisation of the conical singularity in space with time and space coordinates exchanged.  The  $r\to\infty$ conformal boundary of the space is the  torus  $\mathcal{T}$($2\pi\Gamma L,\Gamma\beta_{con}$), which is the torus associated with conical singularities in space with exchanged cycles. This $
\mathrm{AdS_{3}}$ space  with conical defect in time 
is dual to a CFT living on  this torus.\\

\end{itemize}

\section{Modular transformations}
\label{sec3}
One useful concept when dealing with a  2D CFT in the complex torus is that of modular invariance (see e.g. Ref. \cite{DiFrancesco:1997nk,Blumenhagen:2009zz}). The partition function of the theory must be invariant  under modular transformations.
The most important parameter for CFT on torus is the modular parameter $\tau=\omega_2/\omega_1$ where 
$\omega_{1,2}$ are periods of the torus. With the help of this modular parameter $\tau$, various modular transformations are 
defined which together form the modular group {\it{PSL}}(2,$\mathbb{Z}$)\,: $\tau \to (a \tau+b)/(c\tau+d)$ with $ad-bc=1$.
There are three types of modular transformations  \cite{DiFrancesco:1997nk,Blumenhagen:2009zz}: $\mathcal{T}$ : $\tau \rightarrow \tau + 1$, $\mathcal{S}$ : $\tau \rightarrow - \frac{1}{\tau}$, $\mathcal{U}$: $\tau \rightarrow \frac{\tau}{\tau + 1}$.
These transformations generate themselves by composition:
$\mathcal{T=USU ,\hspace{0.3cm} U = TST,\hspace{0.3cm} S= UT}^{-1}\mathcal{U, \hspace{0.3cm} (ST)}^{3}= \mathcal{S}^{2} = 1 $. 

Generally, lattice representation is used to look at modular transformations.
In the lattice representation, torus is identified with vectors  in the complex plane ($\omega_{1,2}$). 
Modular parameter is now given as $\tau = \frac{\omega_{2}}{\omega_{1}} = \tau_{1}+ i\tau_{2}$. 
Considering a unit cell, modular transformations will act on this unit cell
and for simplificity we can choose $\omega_{2}$ = $\tau$ and $\omega_{1}$ = 1.

Of particular relevance for our investigations is the  modular $\mathcal{S}$ transformation. 
It is realized in terms of the action on the complex coordinate on the torus $z= t_E+ix_E$ ($x_E,t_E$ are Euclidean space and time) and on the modular parameter $\tau$ as 
\begin{equation}\label{1a}
 z\to z'  = \frac{z}{\tau}, \quad\quad  \tau\to \tau' = -\frac{1}{\tau}.
\end{equation}

In the context of the  AdS$_3$/CFT$_2$ correspondence $\mathcal{S}$ transformations have been  used focusing on  their action on the modular parameter $\tau$ associated with the boundary tori dual to 3D bulk configurations. It has been shown that the modular parameter $\tau_{AdS}$ of the CFT dual to AdS$_3$ at finite temperature is related to  the modular parameter  $\tau_{BTZ}$  of the CFT dual to the BTZ black hole by the  modular modular $\mathcal{S}$ transformation $\tau_{AdS}= -1/\tau_{BTZ}$ \cite{Aharony:1999ti, Carlip:1994gc}. Later, it has been shown  that  same relation  holds for the modular parameter $\tau_{con}$ of the CFT dual to  AdS$_3$ with space conical singularities:  $\tau_{con}= -1/\tau_{BTZ}$ \cite{Cadoni:2009tk,Cadoni:2010kla}. 

Using the results of the previous section we can now also show  that  the boundary torus dual to AdS$_3$ with  conical singularity along the time direction belongs to the web generated by $\mathcal{S}$ transformations.  In fact, we have seen that the boundary tori associated respectively to conical singularities in space  and conical singularities in time are given respectively  by $\mathcal{T}$($\Gamma\beta_{con},2\pi\Gamma L$) and $\mathcal{T}$($2\pi\Gamma L$, $\Gamma\beta_{con})$. They have their cycles exchanged and are therefore connected by a $\mathcal{S}$ duality transformation.

Having in mind  the classification of the 3D bulk configuration given in Sect. \ref{sec2}, we conclude that all the boundary tori dual to 3D AdS gravity bulk configurations are related one with the other by means of  and $\mathcal{S}$  modular transformation.
This means that compatibility of the AdS$_3$/CFT$_2$ correspondence  with the modular symmetry of the torus requires that with the exception of the vacua, the four types of bulk configurations listed in Sect. \ref{sec2} (thermal AdS, BTZ black hole, conical singularities in space, conical singularities in time) 
 have to be included in the physical spectrum of the theory.  This is a quite non trivial result because conventional wisdom would require  singular geometries to be excluded from the physical spectrum of pure AdS$_3$ gravity.

As noted above, $\mathcal{S}$  modular transformations (\ref{1a}) are generally used only in terms of their action on the modular parameter $\tau$ 
and not in terms of their action on the complex coordinate $z$ of the torus.  This is simply due  to the obvious fact that the partition function for a  CFT on the torus depends on  $\tau$ but not on $z$. 
On the other hand this is not anymore true when one considers the EE of the CFT on the torus as we do in this paper. 
The EE of  a QFT is defined as the von Neumann  Entropy originated by tracing the density matrix over unobservable degrees of freedom localized in a codimension  one region $B$ of the manifold where the QFT lives. In the case of our 2D CFT  this means that we have to assume that only a slices of say length $l$ localized either in the space or time direction of the torus is accessible to measurement. 

One can now easily show that the modular transformation (\ref{1a}) exchanges real and imaginary part of the complex coordinate $z$, thus it can be used 
to exchange slices localized in the space  and time directions. In fact,  
taking for simplicity $\omega_{1} = 1$,  $\tau=\omega_{2}=i\alpha$ ( with $\alpha$ real number) and
a slice localized  in the (Euclidean) time dimension, $z=l$, using Eq. (\ref{1a}) we find 
 \begin{equation}\label{1b}
 z'  =  - \frac{i l}{\alpha}, \quad \quad \tau' = \frac{i}{\alpha}.
\end{equation}
This  means moving from real to imaginary axis, hence using this aspect
of $\mathcal{S}$ transformation we can exchange slices between (Euclidean) space and  and time directions.
In general one could consider the most general case in which slices have both non vanishing real and imaginary part.
For simplicity, in this paper we will limit our considerations to slices localized completely either  in the space or time dimension.

\section{Cylinder approximation and  modular transformations acting on boundary cylinders}
\label{sec4}
As we shall discuss  in the next sections, the EE for 2D CFT in the torus has not an universal form but depends on the actual field content of the theory.
Conversely for  a 2D CFT in the cylinder the EE has an universal form.
For this reason it is very useful to consider  the cylinder approximation of the torus $\mathcal{T}(\b,\a)$ in the limit when one of the two compact dimensions decompactifies. This  cylindric limit  of the torus   $\mathcal{T}(\b,\a)$ can be achieved in two different ways. $(1),\, \b\gg \a$: in this case the time direction of the torus decompactifies  and we have $\mathcal{T}(\b,\a)\to \mathcal{C}(\a)$, where $\mathcal{C}(\a)$ is defined as the cylinder with  noncompact time  direction and compact space direction of length $\a$  ;  $(2),\, \a\gg \b$: in this case is the space direction of the torus that decompactifies  and we have $\mathcal{T}(\beta,\alpha)\to {C}(\b)$, where $C(\b)$ is defined as the cylinder with  noncompact space  direction and compact time direction of length $\b$.

The modular transformations $\mathcal{S}$  described in  the previous section  for the torus  have a natural extension when we work in the cylinder approximation. Applying  Eq. (\ref{1a}) to our torus $\mathcal{T}(\b,\a)$, we see that it is transformed into a torus with modular parameter $\tau'= i\b/\a$, i.e into the torus  $\mathcal{T}(\a,\b)$.  Thus in the limit $\b\gg \a$ the $\mathcal{S}$ transformation acts as $\mathcal{C}(\a)\to C(\a) $. Conversely, in the limit   $\a\gg \b$,  $\mathcal{S}$ acts as as $ C(\b)\to\mathcal{C}(\b)$. Hence,  modular transformations act on the boundary cylinders by exchanging compact with non compact directions. 
By analogy with the usual thermal definition,
where the periodicity  of euclidean time, $\beta$ gives the scale of thermal correlation, i.e the inverse temperature $\b=1/T$, 
we can interpret the periodicity of the  compact space dimension $\alpha$ as giving the scale of  the quantum entanglement correlations. This can be used to define  a sort of "quantum" temperature $\a=1/T_Q$.

 It is important to stress that because the   modular transformation $\mathcal{S}$  exchanges the cycles of torus, it maps different regimes of the CFT. In the cylinder approximation this means that 
the modular map $\mathcal{C}(\a)\to C(\a) $, relates the regime, where $T<<T_Q$ (the regime where quantum correlations dominate) to the  regime $T>>T_Q$, where thermal correlations dominate.  Obviously, for the map $ C(\b)\to\mathcal{C}(\b)$ the opposite is true.

Using the results of section \ref{sec3}, the previous discussion can  be easily extended to the cylindric approximation of 
tori with unobservable regions.
When unobservable regions are present the  $\mathcal{S}$  transformation  not only maps the cylinder $ \mathcal{C}(\b)$ into
$ C(\b)$  but also moves simultaneously  the unobservable region  $B$ from the space to time direction of the cylinder and viceversa. 
Notice that the  $\mathcal{S}$  transformation cannot move regions from compact to noncompact direction of the cylinder. 
An  unobservable region  localized in the compact (non compact) direction already remains in the same direction.

On the other hand, slices can be moved from the compact to the non compact dimension of the cylinder first by using 
$\mathcal{S}$ modular transformation in the torus then taking the infinite size limit of the appropriate direction. Notice that in this case the cylinders are not related by modular transformations.
In the next section we will use  this procedure to move  slices from compact to non compact directions of the cylinder.

\section{Entanglement entropy of boundary CFT in the cylinder }
\label{sec5}
In this section we will compute the EE of the boundary CFT  for  four different positions of slices  and  for the case in which the boundary CFT is defined  on an infinite cylinder. The four different positions are obtained by putting slices in the Euclidean space/time direction and by considering cylinders with compact/noncompact space direction.  We will  consider infinite cylinders instead of tori because on  cylinders  the results for the EE are universal as compared to torus where the results depend on the details of  underlying CFT theory. Later, on Sect. \ref{sec6} we will use these computations to associate a holographic EE to the four gravitational bulk  configuration  described in Sect. \ref{sec2}.

In order to calculate the EE entropy for the CFT on the infinite cylinders  we will use  the following strategy. We will start from the  term in the EE for a free Dirac fermion on the torus of Ref. \cite{Azeyanagi:2007bj},  which  gives in the infinite cylinder approximation the 
leading contribution to the EE entropy for the CFT. Using the modular symmetries   and the universality of the EE for CFT in the cylinder we will derive the EE for the  four different positions of the slice.

Usually, to define entanglement entropy in a QFT we break  the region  where the  degrees of freedom (DOF) live into two disjoint sets $A$ and $B$, then we  
trace  the  density matrix    $\rho$ for the full system over the DOF localized  in the region $B$ to get the reduced density matrix for 
A as $\mathrm{\rho_{A} = Tr_{B}\rho}$. The 
entanglement entropy is then  the Von Neumann entropy: $\mathrm{S_{E} =- Tr(\rho_{A}log\rho_{A})} $ \cite{Calabrese:2004eu,Calabrese:2009qy}.
Specializing to the case of a 2D CFT in the torus, we start  from the result of 
Ref. \cite{Azeyanagi:2007bj} (see also \cite{DiFrancesco:1997nk}), which gives in the  infinite cylinder approximation the universal,  leading contribution to the EE of free Dirac fermions, 
\begin{equation}\label{15}
S_{1} = \frac{c}{3}\ \mathrm{log} \Biggr| \frac{\theta_{1}(z\vert\tau)}{2\pi\eta(\tau)^3}\Biggr|,
\end{equation}

where $z = t_{E} + ix_{E}= it_M + ix_M$ is the slice describing the subsystem $A$   and  $\tau = \beta + i\alpha$   are, respectively,  the complex coordinate and the modular parameter  of the torus,
($t_{E},x_E$ are   euclidean coordinates, $t_{M},x_M$ are  Minkowskian coordinates)\footnote{Notice that in this paper we are using  a definition of the complex coordinate $z$ of the torus different from that  of Ref. \cite{Azeyanagi:2007bj}, where it is defined as $z= x_E+ i t_E$.  }.
Further, $\beta$ is the periodicity for $t_{E}$ and $\alpha$ is the periodicity for $x_{E}$,  $\theta$ is the theta function and $\eta$ is the
Dedekind eta function.

 In  the lattice representation of this torus and in the pictures that follow, we have 
the real  axis along the vertical axis, whereas the imaginary axis is directed along the horizontal axis.
Notice that we are using here the  gereralized form  of the formula given in Ref. \cite{Azeyanagi:2007bj}, where the formula has been used for spacelike slices only. 
Eq. \eqref{15} has a general validity and can be therefore used to compute the entanglement entropy for slices with both space and time components,  just by giving $z$ the desired value.
\begin{itemize}
\item
{\bf Slice\ along \ compact  \ time\ dimension}

 We take the slice along time axis only, that is we consider a slice of the time coordinate of measure $\lt$, 
$z=l_t$ and $\beta=1$,  $\tau = i\alpha$.
We now have $q$ and $y$ as required by the theta function, 
\begin{align*}
q = e^{2\pi i\tau} = e^{-2\pi\alpha},\quad 
y = e^{2\pi iz} = e^{2\pi i \lt}.
\end{align*}
Using the above values for $q$ and $y$, we have the form for $S_{1}$ as:
\begin{equation}\label{16}
S_{1} = \frac{c}{3}\mathrm{log}\Biggr|\frac{1}{\pi}\sin(\pi \lt)\prod_{m=1}^{\infty}\frac{(1-yq^m)(1-y^{-1}q^{m})}{(1-q^m)^2}
\Biggr|.
\end{equation}\\
Pictorially this will look like:
\begin{center}
\begin{tikzpicture}
\node[cylinder,draw,rotate =180, minimum height = 2cm,minimum width = 1cm]{};
\draw(-1.3,0)--(-1,0) (-1.3,0.1)--(-1,0.1); 
\node()at (-1.6,0.05){$\lt$} ;
\node()at (-1.7,-0.3){$(\beta)$};
\node()at (0,-1.2){$(i\alpha)$};
\node()at (2.4,0){where the symbol};
\draw(4,0)--(4.4,0) (4,0.1)--(4.4,0.1); 
\node()at (6.4,0){stands for a slice.};
\end{tikzpicture}
 \end{center}
Now, we take the Eq. \eqref{16} in the infinite spatial size  limit $\alpha>>1$ which corresponds to $T>>T_Q$ (where $T_Q$ is the "quantum" temperature defined in the previous section. 
This gives us the entanglement entropy  on the infinite cylinder $C(\beta)$:
\begin{equation}\label{17}
S^{(tc)} = \frac{c}{3}\mathrm{log}\Biggr|\frac{\beta}{a\pi}\sin\biggl(\frac{\pi \lt}{\beta}\Biggr)\Biggr|, 
\end{equation}
where we have reinstated the periodicity $\beta$ of $t_E$ and   $a$ is an UV cutoff.
 This equation has general  validity  and gives the EE entropy for slices localized along a compact time dimension.
\item
{\bf Slice\ along \ noncompact   \ space \ dimension} 

 We now apply modular transformation $\mathcal{S}$ on $z$ and $\tau$ to get $z'$ and $\tau'$
which is:
\begin{equation}\label{19}
z' = \frac{z}{\tau} = - \frac{i\ls}{\alpha}, \quad 
\tau' =  - \frac{1}{\tau} = \frac{i}{\alpha}, 
\end{equation}
where we have renamed $\lt$ into $\ls$ to stress  its different character (measure of space slice  instead of time slice). 
As a result, we now have $q$ and $y$ as:
\begin{equation}\label{20}
\begin{split}
q = e^{2\pi i\tau} = e^{-2\pi/\alpha}\ \, \quad
y = e^{2\pi iz} = e^{2\pi \ls/\alpha} .
\end{split}
\end{equation}
Modular transformation $\mathcal{S}$ has taken the slice from real to imaginary axis and pictorially this looks like:
\begin{center}
\begin{tikzpicture}
\node[cylinder,draw,rotate =180, minimum height = 2cm,minimum width = 1cm]{};
\draw(0,-0.3)--(0,-0.6) (0.1,-0.3)--(0.1,-0.6); 
\node()at (-1.5,0){$(\alpha)$};
\node()at (0,-0.8){$i \ls$};
\node()at (0,-1.2){$(i\beta)$};
\end{tikzpicture}
\end{center}
Using the modular transformation $\mathcal{S}$, we now have the form for $S_{1}$ as :
\begin{equation}\label{21}
S_1  = \frac{c}{3}\mathrm{log}\Biggr|\frac{\alpha}{\pi} e^{-\textstyle{\frac{\pi l^2}{\alpha}}}\sinh(\frac{\pi \ls}{\alpha})
\prod_{m=1}^{\infty}\frac{(1-yq^m)(1-y^{-1}q^{m})}{(1-q^m)^2}\Biggr|. 
\end{equation}
Now, we take Eq. \eqref{21} in the infinite space size, $\alpha<<1$ limit, implying $T>>T_Q$,
which gives us the entanglement entropy of the CFT on the infinite cylinder $C(\a)$ with   slice localized in the spatial, noncompact, dimension:
\begin{equation}\label{22}
S^{(snc)}  = \frac{c}{3}\mathrm{log}\Biggr|\frac{\alpha}{a\pi}\sinh\biggl(\frac{\pi \ls}{\alpha}\biggr)\Biggr| .
\end{equation}

\item
{\bf Slice\ along \ compact   \ space \ dimension}

 \vspace{0.2cm} 
 Next, we    apply the  modular transformation $\mathcal{S}$ on the cylinder described in section
 \ref{sec4}, to the cylinder $C(\beta)$ with slice in the  compact time dimensions 
 to obtain the cylinder $\cal{C}(\beta)$ with slice in the compact space dimension.
Pictorially this looks like:
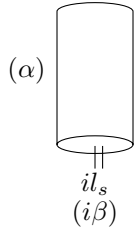
\begin{center}
\begin{tikzpicture}
\node[cylinder,draw,rotate =270, minimum height = 2cm,minimum width = 1cm]{};
\draw(0,-1)--(0,-1.3) (0.1,-1)--(0.1,-1.3); 
\node()at (0,-1.5){$i\ls$};
\node()at (0,-1.9){$(i\beta)$};
\node()at (-0.9,0){$(\alpha)$};
\end{tikzpicture}
\end{center}
where, as usual, we have renamed $\lt$ into $\ls$. This cylinder approximation holds for $\a\gg \b$, i.e for $T\ll T_Q$.
Eq. (\ref{17}) does not change under the modular transformation $C(\beta)\to \cal{C}(\beta)$ and, correspondingly, we get  the EE for a CFT on the $\cal{C}(\b)$ cylinder with slice $\ls$along the compact space dimension.

\begin{equation}\label{17a}
S^{(sc)} = \frac{c}{3}\mathrm{log}\Biggr|\frac{\beta}{a\pi}\sin\Biggl(\frac{\pi \ls}{\beta}\Biggr)\Biggr|. 
\end{equation}

\item
{\bf Slice\ along \ noncompact   \ time \ dimension}

To get the EE for slice along non compact time dimensions we use the modular transformation $\mathcal{S}$ on the cylinder described in section  \ref{sec4}, to the cylinder $C(\a)$ with slice in the non compact space dimensions   to obtain the cylinder $\cal{C}(\a)$ with slice in the non compact time dimension.
Pictorially, after renaming $\ls$ into $\lt$ this looks like:
 
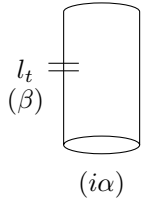
\begin{center}
\begin{tikzpicture}
\node[cylinder,draw,rotate =270, minimum height = 2cm,minimum width = 1cm]{};
\draw(-0.3,0)--(-0.7,0) (-0.3,0.1)--(-0.7,0.1); 
\node()at (0,-1.5){$(i\alpha)$};
\node()at (-1.0,0){$\lt$};
\node()at (-1.0,-0.4){$(\beta)$};
\end{tikzpicture}
\end{center}
The modular transformation $C(\a)\to \cal{C}(\a)$ leaves the EE (\ref{22}) invariant. This allows us to write  the EE 
of a CFT on the $\cal{C}(\a)$ cylinder with slice $\lt$ along the noncompact time dimension: 
\begin{equation}\label{22a}
S^{(tnc)}  = \frac{c}{3}\mathrm{log}\Biggr|\frac{\alpha}{a \pi}\sinh\Biggl(\frac{\pi \lt}{\alpha}\Biggr)\Biggr| .
\end{equation}
This cylinder approximation holds for $\b\gg \a$, i.e for $T\ll T_Q$.
\end{itemize}

The four expressions (\ref{17}),(\ref{17a}) and (\ref{22}),(\ref{22a})  for the EE of a CFT$_2$ on the cylinder we have derived in this section are universal and  well-known \cite{Holzhey:1994we,Calabrese:2004eu,Calabrese:2009qy}. However we want to stress here that their physical meaning and regime of validity is slightly  different from what is usually considered. Eq. (\ref{17}) and (\ref{22}) give the  contribution of, respectively,   time and space entanglement  at large temperature when thermal correlations dominate, whereas Eq. (\ref{17a}) and (\ref{22a})  give the same information but at small $T$, when quantum correlations dominate.

\subsection{Planar approximation}

The planar approximation describing a CFT in the plane can be obtained by considering in Eqs. (\ref{17}),(\ref{17a}) 
the limit $l_{t,s}<<\beta$ and in Eqs. (\ref{22}),(\ref{22a}) the limit $l_{t,s}<<\a$.
We get the same expression for all four cases,  
giving the EE entropy for a CFT a zero temperature, 
non compact space direction and  with a space (time) slice of length $l_s$ ($l_t$):

\begin{equation}\label{pl}
S^{(pl)}  = \frac{c}{3}\mathrm{log}\Biggr|\frac{l_{s,t}}{a}\Biggr| .
\end{equation}

\section{Holographic entanglement entropy of bulk gravitational configuration}
\lb{sec6}
In this section we will show how, starting from the results of Sect. \ref{sec5},   one can assign   an holographic EE to the 3D bulk gravity configurations listed in Sect. \ref{sec2}. Despite some progress, the problem of reducing  the EE of bulk gravitational configuration to dual boundary CFT calculations is not solved \cite{Solodukhin:2006xv,Azeyanagi:2007bj,Cadoni:2007vf,Cadoni:2009tk,Cadoni:2010kla,Nishioka:2009un,
Takayanagi:2012kg,Chaturvedi:2016kbk,Chaturvedi:2016rft,Singh:2016bld}.
Our procedure is based on the AdS/CFT correspondence, which implies that in the large $c$ limit we can use the dual CFT to describe 3D bulk configurations. The use of the dual CFT for calculating the EE has several advantages with respect to other non-holographic approaches, in which the EE entropy of  gravity configurations, like black holes,  is defined in terms of the EE of  quantum states of matter fields in the  given, classical, gravitational background \cite{tHooft:1984kcu,Mann:1996ze}.
Typically, this bulk  calculations are not universal, but depend on the  number of matter fields and on
the UV cutoff \cite{tHooft:1984kcu,Mann:1996ze}.
Conversely,  our approach does not suffer of these shortcoming.
In a sense the number of  matter fields is fixed by  the central charge $c$ (\ref{cc}) of the dual CFT and 
the UV cutoff $a$ appearing in  the previous sections is determined by the UV/IR relation, in terms of an infrared cutoff \cite{Cadoni:2009tk,Cadoni:2010kla}
\be\lb{irc}
\Lambda= 4\pi^2L^2/a,
\ee
 on the radial coordinate of AdS$_3$ .

The AdS/CFT correspondence  can be used in two different ways to determine the EE associated with 3D gravitational 
configurations. The first way, uses the results of the previous section, i.e   the fact that the 
EE of a CFT in the cylinder has an universal form. This will be discussed in this section. 
The second way will be discussed in the next section and uses the Ryu-Takayanagi prescription \cite{Ryu:2006bv,Ryu:2006ef}, differential entropy \cite{Balasubramanian:2014sra} and complexified geodesics \cite{Balasubramanian:2012tu}. 

In view of the results of Sect. \ref{sec5}, one could  find the holographic EE of the bulk gravitational configuration by  simply applying the appropriate equations in (\ref{17}),(\ref{22}),(\ref{17a}),(\ref{22a}) to the boundary of the 3D gravity solution listed in Sect. \ref{sec2}. This will give universal expressions for the EE entropy in the regime in which  boundary tori can be approximated by cylinders, i.e in the large and small temperature regime.
However, this cannot be done so simply because   Eqs.  (\ref{17}),(\ref{22}),(\ref{17a}),(\ref{22a}) depend on an arbitrary boundary parameter, the slice lengths
$l_t,l_s$, giving the measure of the observable region $A$ for the CFT.
Because correlations of the bulk theory are codified in a highly non local way in the boundary CFT there in no natural way to identify $l_t$ and $l_s$ in terms of bulk parameters characterising 3D bulk geometries.

Using arguments borrowed from the IR/UV relation in the AdS/CFT correspondence,   in Ref. \cite{Cadoni:2009tk,Cadoni:2010kla}  an  identification  $l_s=2\pi L$ was proposed for the case of the BTZ black hole. The proposal had a  conjectural status. Let us now demonstrate it using geometric arguments. We also extend its validity to all the gravity solutions listed in  Sect. \ref{sec2}.\\
\begin{itemize}
\item
{\bf Entanglement\ entropy \ of \ AdS$_3$ \ with  \ conical   \ singularity  \ in \ space}

Let us start from the 3D bulk solution  (\ref{12}) describing AdS$_3$ with conical singularities along the  $\phi$ direction.
The dual boundary CFT lives in the boundary  torus ${\cal{T}}(2\pi L, 2\pi r_+)$. The key observation is that for $r_+<L$ we can see the conical singularity as originated by an observable  slice of length $2\pi r_+$ in the spatial cycle of total length $2\pi L$ of the boundary torus  ${\cal{T}}(2\pi r_+, 2\pi L)$. Notice that this torus is related to the original torus by a modular transformation (it  has exchanged cycles). Moreover,    putting $r_+=0$  the conical singularity in the  bulk  disappears. For $r_+=0$ we get  the physical vacuum  of the theory (in the bulk AdS$_3$ in  Poincar$\mathrm{\acute{e}}$ patch, in the  boundary a $T=0$ CFT). 

By rescaling  the lengths by a factor $L/r_+$ we get the torus 
${\cal{T}}^{cs}( 2\pi L,\beta_{con},)$ and  for $r_+<<L$ the cylinder    ${\cal{C}}^{cs}(\b_{con})$ with slice length $l_s=2\pi L$.
The holographic EE  for AdS$_3$ with a conical singularity in space is therefore given by Eq. (\ref{17a})
with $\b=\b_{con}=2\pi L^2/r_+$ and $l_s=2\pi L$:
\begin{equation}\label{17ab}
S^{(cs)} = \frac{c}{3}\mathrm{log}\Biggr|\frac{\beta_{con}}{a\pi}\sin \Biggl(\frac{2\pi^2 L)}{\beta_{con}}\Biggr)\Biggr|.
\end{equation}

Notice that as expected this approximation holds for small temperatures, i.e for $T=1/\beta_{con}<<T_Q=1/(2\pi L)$.\\

\item
{\bf Entanglement\ entropy \ of \ AdS$_3$ \ with  \ conical   \ singularity  \ in \ time} 

The boundary torus associated to AdS$_3$ with conical singularity in time is obtained by applying the ${\cal{S}}$ transformation to ${\cal{T}}^{cs}( 2\pi L,\beta_{con},)$ with slice length $l_s=2\pi L$.
We get  ${\cal{T}}^{ct}(\beta_{con}, 2\pi L,)$  with slice length $l_t=2\pi L$. 
Correspondingly, for $L>>r_+$ the above constructed boundary cylinder  ${\cal{C}}^{cs}(\b_{con})$ with spatial slice of length $l_s=2\pi L$,  associated with AdS$_3$ with  conical singularity in space, is mapped by the ${\cal{S}}$  modular transformation  in to the boundary cylinder 
$C^{ct}(\b_{con})$ with  slice length $l_t=2\pi L$,  associated with AdS$_3$ with  conical singularity in time.
Thus, we get from Eq. (\ref{17})  the EE of AdS$_3$ with a conical singularity in time
\begin{equation}\label{17c}
S^{(ct)} = \frac{c}{3}\mathrm{log}\Biggr|\frac{\beta_{con}}{a\pi}\sin\Biggl(\frac{2\pi^2 L_{t}}{\beta_{con}}\Biggr)\Biggr|,
\end{equation}
where we have used $L_{t}\equiv L$ to stress the fact that now $L_t$ is the length of an {\sl euclidean time } slice and the formula holds in the regime $T>>T_Q$. Notice that although  in Euclidean space $L_t=L$ and   the holographic EE for AdS$_3$ with conical singularities in space and time are the same,  the meaning of  Eqs. (\ref{17ab}) and (\ref{17c}) is rather different. Whereas
Eq. (\ref{17ab}) gives the holographic EE entropy of conical singularities in space in the regime where quantum correlations dominate, Eq. (\ref{17c}) gives the holographic EE entropy of conical singularities in time  in the regime where thermal correlations dominate.

\item
{\bf Entanglement\ entropy \ of \ the  \ BTZ   \ black    \ hole  }

It is known that the boundary torus  for AdS$_3$ with a conical singularity in space ${\cal{T}}^{cs}(2\pi L, \b_{con})$  
and  the boundary torus ${\cal{T}}^{BTZ}( \b_{H},2\pi L)$ associated with the BTZ black hole are related   by a ${\cal{S}}$  
modular transformation \cite{Cadoni:2009tk,Cadoni:2010kla}. Working in the, $r_+>> L$, cylinder approximation and using the results of 
Sect. \ref{sec5}   allows us to build the cylinder map ${\cal{C}}^{cs}(\b_{con})\to   C^{BTZ}(\b_{H})$, where the two cylinders have slice of length $2\pi L$ respectively along the compact space and non-compact space directions. 
Correspondingly, Eq. (\ref{22}) with $\a=\b_H$ and $l_s=2\pi L$ will gives us the EE entropy of the BTZ black hole
\begin{equation}\label{22b}
S^{(BTZ)}  = \frac{c}{3}\mathrm{log}\Biggr|\frac{\b_H}{a \pi}\sinh\Biggl(\frac{2\pi^2 L}{\b_H}\Biggr)\Biggr|,
\end{equation}
which holds in the regime $T>>T_Q$ and  fully coincides with the result of \cite{Cadoni:2009tk,Cadoni:2010kla}.

We can now close the web of modular transformations by mapping the cylinder  $C^{BTZ}(\b_{H}) $ with slice in the non compact space direction into the ${\cal{C}}^{BTZ}(\b_{H})$ with slice along the noncompact time dimension. Correspondingly Eq. (\ref{22a})  gives
\begin{equation}\label{22ac}
S^{(BTZ)}_{ET}  = \frac{c}{3}\mathrm{log}\Biggr|\frac{\b_H}{a \pi}\sinh(\frac{2\pi^2 L_t}{\b_H})\Biggr|,
\end{equation}
giving the holographic EE entropy of the BTZ black hole in terms of Euclidean time correlations. Eq. (\ref{22ac}) holds in the region $T<<T_Q$.
 
Notice that also here  we have used $L_{t}\equiv L$  to represent the length of an {\sl euclidean time } slice.
 Considerations similar to those concerning the relation between Eqs. (\ref{17ab}) and (\ref{17c}) 
apply also to Eqs. (\ref{22b}) and (\ref{22ac}). Eq. (\ref{22b}) gives the holographic EE entropy of the BTZ black hole
in the regime where thermal correlations dominate and  in terms spatial correlations. On the other hand   Eq. (\ref{22ac}) gives the same quantity but in terms of time correlations  and  in the regime where quantum correlations dominate.

\item
{\bf Entanglement\ entropy \ of \ the  \ AdS$_3$   \ vacuum     }

The leading term in the $ r_+<<L$ limit    of  Eqs. (\ref{17ab}) and \ref{22ac}) give the entanglement entropy of AdS$_3$ vacuum, which can be also  expressed in terms of the IR cutoff
$\Lambda$ of Eq. (\ref{irc}), 
\begin{equation}\label{pl1}
S^{vac}  =\frac{c}{3}\mathrm{log}\Biggr(\frac{2\pi L}{a}\Biggr)= \frac{c}{3}\mathrm{log}\Biggr(\frac{\Lambda}{L}\Biggr).
\end{equation}
This equation can be obtained directly from the expression (\ref{pl})  giving the EE of the CFT in the plane setting $l_{s,t}= 2\pi L$ and then using (\ref{irc}).
Notice also that the EE of the AdS$_3$ vacuum cannot be obtained from Eqs. (\ref{17c}) or (\ref{22b}) because they are large temperature ($r_{+}>>L$) expansions.
\end{itemize}

Let us now briefly summarise the main results of this section. 
Pure 3D AdS gravity allows for  two classes of solutions corresponding to  singularities 
shielded by event horizons and naked (conical) singularities produced by a pointlike mass.  
In the cylinder approximation, i.e in  the  large or small temperature limit, 
the  holographic EE $S_E$ associated with these two classes of configurations falls also in two universality classes. 
Spacetimes with event horizons are always  described by  CFTs living in cylinders where regions of 
the non compact dimension are not accessible to observation and the corresponding EEs (\ref{22b}),(\ref{22ac}) 
have an exponential  behaviour.   Spacetimes with conical singularities
are always  described by  CFTs living in cylinders where regions of the  compact dimension are not accessible to observation and the corresponding EEs (\ref{17ab}),(\ref{17c}) have a periodic  behaviour.

\section[Holographic entanglement entropy from gravitational tools]
{Holographic entanglement entropy  from gravitational tools}
\label{sec7}
In the framework of AdS/CFT correspondence, entanglement entropy of the boundary CFT  can also be calculated using gravitational tools.
In order to do so, one may use the holographic entanglement entropy formula as proposed by Ryu-Takayanagi \cite{Ryu:2006bv,Ryu:2006ef}.
As per this formalism, considering a $d$-dimensional CFT, the entanglement entropy for a region $A$ having a $d-2$ dimensional 
boundary $\partial A$ is given by the minimal area of the $d-1$ dimensional surface $\gamma_{A}$ 
in $\mathrm{AdS}_{d+1}$ whose boundary $\partial\gamma_{A}$ coincides with $\partial A$:
\begin{equation}\label{36}
S_{A} = \frac{\mathrm{Area}(\gamma_{A})}{4G_{N}^{d+1}}.
\end{equation}
In this section we will show how the RT prescription  and its extensions \cite{Balasubramanian:2012tu,Balasubramanian:2014sra} can be used to assign  a holographic EE to the 3D bulk gravity  configurations and to reproduce our results in  Sect. \ref{sec6}.

\begin{itemize}
\item
{\bf BTZ  black hole}

Entanglement Entropy for  the 2D CFT dual to  BTZ black hole in the bulk
has
been calculated, using the  Ryu-Takayanagi formalism, in  several papers (see e.g \cite{Azeyanagi:2007bj}). In this case the area 
$\mathrm{Area}(\gamma_{A})$ is  equal to the minimal geodesic length between the two endpoints of  1-dim slice $A$ at constant time of length $l_s$ and we get back our equation (\ref{22}) with $\b=\b_H$, where  $\beta_H$ is the inverse Hawking temperature of  the black hole.  

As it has been observed in Ref. \cite{Azeyanagi:2007bj} the EE (\ref{36}) gives also information about the holographic EE of the bulk configuration (the black hole)  on which the Eq. (\ref{36}) is evaluated. In particular,  when $\ls$ becomes large the region $A$ covers most of the constant time-conformal boundary of the BTZ black hole. In this case, the minimal length  geodesic  consists of two disconnected pieces, one of finite length, wrapping around the black hole and an  other, infinitely long of length $\propto \frac{c}{3}\log(\epsilon/a)$ \cite{Azeyanagi:2007bj},  giving essentially the EE of the AdS$_3$ vacuum.  The finite term has to be identified with the holographic EE of the BTZ black hole. In the cylinder approximation  $\beta<<2\pi L$ we have to consider $\phi$ living in the universal covering  of the circle $S^1$. Thus,  the configuration when the geodesic wraps around the black hole corresponds to $ l_s=2\pi L$, which inserted into Eq. (\ref{22}) gives exactly our  holographic EE for the BTZ black hole of Eq. (\ref{22b}).

\item
{\bf Conical\ singularity\ in\ space}

Entanglement entropy for  the 2D CFT dual to  a spatial  conical singularity in the bulk  has been calculated in Ref. \cite{Balasubramanian:2014sra}:
\begin{equation}\label{38}
S_{con}= \frac{c}{3}\mathrm{log}\Biggr|\frac{R}{\pi(\Gamma\mu)}\sin(\Gamma\phi)\Biggr|,
\end{equation}
where in the above equation, $R$ is the total size of the system, $\Gamma\phi$ is the 
subsystem size in angular coordinates, $\mu$ is the UV cutoff.

Also in this case we expect Eq. (\ref{38}), which gives the EE of the 2D CFT living in the boundary,  to give the information about the holographic EE of the bulk configuration. 
This information can be easily extracted by recalling that, as pointed out in Sect. \ref{sec6}, 
for $r_+<L$, we can see the conical singularity as originated by a slice of length $2\pi r_+$  in the spatial cycle of length 
$2\pi L$.

This means that in Eq. (\ref{38}) we have to  take $R= 2\pi L$ and $\Gamma\phi = \pi r_{+}/L$. 
By appropriate identification of the UV cutoffs,   $\Gamma\mu = a r_{+}/L$ and using the definition  (\ref{bcon}) of  $\beta_{con}$ 
we reproduce exactly our  Eq. \eqref{17ab}.
\item
{\bf Conical\ singularity\ in\ time}

Since we are considering now timelike separated points on the boundary - and it is known that 
timelike geodesics in AdS do not reach the boundary-  to calculate entanglement entropy for 
conical singularity in time, we use the  complexified geodesic approach 
of Ref. \cite{Balasubramanian:2012tu}.
We therefore  use the complexified, renormalized,  geodesic length for   points separated  by the interval $\Delta t_E$, as calculated  in  in \cite{Balasubramanian:2012tu} for the BTZ black hole background
\begin{equation}\label{40a}
\mathcal{L}  = 
2\ \mathrm{ln} \bigg[ \frac{2L}{r_+}\sin(\frac{r_+ \Delta t_E}{2L^2})\bigg]
\end{equation}
where $t_E$ is the Euclidean time with periodicity  $2\pi$. 

Formula \eqref{40a} holds not only for a regular  Euclidean manifold without conical singularities but after a  rescaling of the $t_E$ also for a manifold with a conical singularity.
Similarly to AdS$_3$ with conical singularities in the space direction,  the case of conical singularity in the 
time direction can be described by considering it as described by  time-slice of 
length $\Delta t_E= 2\pi L$. Using this relation and reinstating 
an appropriate UV cutoff $a$ Eq. (\ref{40a}) gives
our result of Eq. \eqref{17c}.

\end{itemize}

\section{ Leading term in the large/small radius expansion of the entanglement entropy and area versus volume scaling }
\lb{sec8}
In Sect. \ref{sec5} we have shown that in the  large/small radius limit the  holographic EE of 3D gravity solutions falls in two classes. In this section we  compute the leading terms  of the large/small radius behaviour and, when possible,  compare it with the thermal entropy of the system.

Let us first consider the BTZ black hole, i.e Eqs. (\ref{22b}) and (\ref{22ac}). The cylinder approximation is valid for $r_+/L>>1$.
Expanding for $r_+/L>>1$ both Eqs. (\ref{22b}) and (\ref{22ac}) give
\beq\lb{jkl}
\Delta S\equiv S^{BTZ}-S^{vac}=
\frac{\pi r_{+}}{2 G_{3}}- \frac{L}{2G_{3}}\ln \frac{\pi
r_{+}}{L}+ O(1)= S_{BH}- \frac{L}{2G_{3}}\ln S_{BH}+ O(1).
\eeq
where we have subtracted the contribution from the AdS vacuum $S^{vac}$ given by Eq. (\ref{pl1}) and $S_{BH}$ is the thermal Bekenstein-Hawking entropy of the BTZ black hole.

Eq. (\ref{jkl}) has a different interpretation when referred to Eq. (\ref{22b}) or to (\ref{22ac}). 
In the first case the $r_+/L>>1$ regime represents a large $T$ behaviour and Eq. (\ref{jkl}) is what one naturally  expects.  In this regime  thermal correlations dominate over  quantum spatial correlations and the leading term in the  EE is just a measure of thermal entropy.
Also the subleading $ \ln S_{BH}$ term   is rather expected \cite{Mann:1996ze,Singh:2011gd,Sachan:2014hna}. From the bulk point of view, the leading term in the large $T$ expansion of the EE is {\sl positive},  scales  as an area and reproduces correctly the area law for black holes.
This means that, as expected,   presence of an horizon in the bulk enhances the EE  of the vacuum and has an holographic character.

On the other hand, from   the point of view of the boundary CFT the leading term in Eq. (\ref{jkl}) has the expected  extensive,  thermal, character. We can easily see from Eq. (\ref{22}) that the regularized, large $T$ leading   term of the  EE for a CFT on the cylinder is given by
\beq\lb{mlk}  
S^{snc}-S^{vac}= \frac{c}{3} \frac{\pi l_s}{\beta}.
\eeq
With the  identification $\ls= 2\pi L$,  Eq. (\ref{mlk}) agrees, as expected from the AdS/CFT correspondence,  with the extensive Gibbs entropy $S_{Gibbs}= (2/3) c\pi^2 L T$ of  a 2D CFT with a  typical spatial scale $L$ and reproduces, after setting $\b=\b_H$, the Bekenstein-Hawking entropy of the BTZ black hole.

The interpretation of Eq. (\ref{jkl}) when referred to  Eq.  (\ref{22ac}), i.e to  the case of time slices, is much more involved.
It will be discussed in the next section. 

Let us now consider conical singularities, i.e Eqs. (\ref{17ab}) and (\ref{17c}).
In this case the the cylinder approximation is valid for $r_+/L<<1$, i.e well below the Hawking-Page phase transition, where thermal 
AdS$_3$ is energetically preferred. However, in this paper we are not interested  in stability questions and will just consider the contribution  coming from conical singularities. Also in this case the discussion about  time slices and the related Eq. (\ref{17c}) is more involved  and it will be postponed to the next section.

Expanding Eqs. (\ref{17ab})  we get at leading order  in $r_+/L$

\begin{equation}\label{39}
\Delta S = -\frac{c \,\pi^2}{18} \Gamma^{2}= -\frac{\pi^2}{12 G_3}\frac{ r_{+}^2}{L}.
\end{equation}

We see that from the 3D bulk point of view in the small $T$ limit the holographic EE of conical singularities is {\sl negative} and  scales as a {\sl  volume}. 
The $r_+/L<<1$ regime has the meaning of a small $T$ expansion, therefore in this regime spatial quantum correlations dominate over thermal correlations. From the point of view of the boundary CFT we have a super-extensive scaling behaviour. 
In fact expanding  Eq. (\ref{17a}) we get at leading order
\begin{equation}\label{40}
\Delta S = -\frac{c}{18}\biggr(\frac{\pi l_s}{\b}\biggr)^{2}= -\frac{c}{18} \pi ^2 l_s^2T_{Q}^2, 
\end{equation}
where $T_Q=1/\b$ is the "quantum" temperature introduced in  Sect. \ref{sec4}.
The negative sign in Eqs. (\ref{39}) and (\ref{40}) means physically  that insertion of a point mass not shielded by an event horizon in the AdS$_3$ vacuum  corresponds to a reduction  of the EE  of the vacuum and gives  a volume contribution. This entanglement reduction due to matter has been also observed in a  4D de Sitter back ground in  Ref \cite {Verlinde:2016toy}.
Moreover, a volume-law scaling of the EE has been found in the subsystem-small-size regime   of a class of non-local field  theories  \cite{Shiba:2013jja}.

\section{Entanglement  entropy in Minkowski space}
\lb{sec9}
In the previous sections we have calculated  the holographic EE considering a spacetime with Euclidean signature.
Now we want to discuss what happens when we consider it in Minkowski spacetime.
Formally, the transition  from  Euclidean to Minkowski space  can be accomplished just by considering  the Wick rotation
$t_E=it$. Obviously, the physical interpretation  of the Wick rotation is far from being trivial.

As long as we take   the subsystem $A$ as a spatial slice, the Wick rotation has no effect on our calculations of the EE.
This means the  Eqs. (\ref{22}),  (\ref{17a})  and correspondingly Eqs. (\ref{17ab}), (\ref{22b}) do not change when passing to 
Minkowski space. Physically, this is a consequence of the fact that they are obtained from correlations in equilibrium states of a thermal QFT.
On the other hand, this is not anymore true, when we consider the EE relative to time slices,  i.e Eqs. 
(\ref{17}),  (\ref{22a})  and correspondingly Eqs. (\ref{17c}), (\ref{22ac}). In the euclidean space, where time $t_E$ and space 
variables $x_E$ are  treated on the same  footing, these equations are simply related to 
(\ref{22}),  (\ref{17a}), (\ref{17ab}), (\ref{22b}) by  simply renaming  the variables $t_E\leftrightarrow x_E$. 
The Wick rotation $l_t\to il_t$ (and correspondingly 
$L_t\to iL_t$) changes   the nature of these  equations. Periodic behaviour in the Euclidean becomes  exponential
in the Minkowskian. This can be applied to Eqs. (\ref{17}) and  (\ref{17c}), which hold in the high temperature regime.  
Physically, this means that after  Wick rotation Eqs. (\ref{17}) and  (\ref{17c})  
describe  correlations in  non equilibrium states of a thermal QFT in the regime dominated by thermal correlations. 
This interpretation of EE for timelike slices in terms of    
nonequilibrium correlations is very similar to the  two-point functions calculations  for timelike intervals of 
Ref. \cite{Balasubramanian:2012tu}.

On the other hand, the Wick rotation  applied  to Eqs. (\ref{22a}), (\ref{22ac}) transforms the exponential behaviour into periodic one. We do not have a clear interpretation of Eqs. (\ref{22a}), (\ref{22ac}) in  Minkowski space. This is because they 
hold in the regime $T<<T_Q$ where quantum correlations  dominate and a thermal interpretation is not appropriate. 
It is likely that the Minkowskian version the entanglement entropies (\ref{22a}), (\ref{22ac}) 
should involve  some notion of "entanglement  in time", for instance like that proposed 
in Ref.  \cite{Olson:2010jy,Olson:2011bq}. This point deserves further investigations.

\section[Causality aspects of holographic entanglement entropy]
{Causality aspects of holographic entanglement entropy}
\label{sec10}

When entanglement entropy for a CFT is calculated using the replica trick, we are assured that 
locality and 
causality are preserved in the boundary  theory. On the other hand, when one uses the Ryu-Takayangi formalism, it is not completely clear how the minimal surfaces
capture  bulk information and express it in terms of boundary spatial entanglement. In \cite{Hubeny:2012wa} the authors have tried to answer to this question 
by constructing the causal holographic information. 
The construction, 
although  similar to the Ryu-Takayanagi  method, is indeed quite different from it. This causal holographic information is 
the proper area of the causal information surface, just like the entanglement entropy is given by the area of the minimal 
surface. Among the cases where the spatial entanglement entropy matches the causal holographic information are  
the Global $\mathrm{AdS_3}$ and BTZ black hole giving support to the fact that at least in these cases the entanglement entropy
given by Ryu-Takayanagi formula is causal.

Combining this causal holographic information associated with a hole in AdS,
the authors in \cite{Balasubramanian:2013lsa} have proposed the concept of  differential entropy which has been used in \cite{Balasubramanian:2014sra} to calculate
the entanglement entropy for conical singularity in space. Differential entropy also helps us to understand the causal
facets of entanglement entropy related to conical singularity in space. This implies   that our holographic derivation of the EE for the BTZ black and for 3D conical singularities in space, i.e Eqs.  (\ref{17ab}), (\ref{22b})   captures in a causal way  bulk information.

These causality aspects become even more involved when we consider  temporal entanglement entropy, i.e our Eqs. (\ref{17c}), (\ref{22ac}).  
In Ref. \cite{Arefeva:2016lnk} the authors do mention about the questionable
aspect of causality related to the complexified geodesics. These complexified geodesics do not guarantee any causal connection
between the boundary and the bulk and presently can only be taken as a mathematical tool to calculate holographic
temporal entanglement entropy.

 Also one can use the image prescription given by \cite{Arefeva:2016lnk} to calculate temporal 
holographic entanglement entropy but in this case also the image prescription is adjusted to take care of causality.
In \cite{Balasubramanian:2014sra} the concept of entwinement has been suggested to probe physics which is not captured by minimal geodesics. One possible application of it  can be to probe black hole interior. It is likely that the temporal entanglement entropy can be considered as a manifestation of entwinement 
where the internal degree of freedom in this case may come from the embedding space of AdS. This is also a point that deserves further investigation.

\section{Conclusion and discussion}
\lb{sec11}
In this paper we have investigated  the signature of the presence of horizons and localized matter in the holographic entanglement entropy in the framework of AdS$_3$ gravity. Our main result is that bulk black hole  excitations of the AdS$_3$ Poincar\'e vacuum respect the area law and enhance the entanglement  of the vacuum. Conversely,  localized matter in the bulk generating conical singularities,  gives a negative contribution to the entanglement  of the vacuum and a contribution scaling as a bulk volume. This negative, bulk volume term corresponds for the boundary CFT to  a super-extensive term.

Using the modular symmetries  and working in the regime in which  boundary tori, 
where the dual 2D CFT lives,  can be approximated by  cylinders we have been able
to  describe black holes and conical singularities  in terms of boundary cylinders with unobservable regions in  the Euclidean space or 
time   directions. This allowed us to  compute  universal expressions for the holographic EE  of the 3D bulk configurations. 
We have shown in  this way that the presence  of black hole excitations in the bulk are encoded in the holographic EE entropy
in terms of presence of unobservable regions in the non compact direction of the cylinder and by the  related  
exponential  behaviour of  the EE. 
Conversely, conical singularities are described by the presence of unobservable regions along  the compact direction of the cylinder and by the 
related  periodic    behaviour of  the EE.

There are several critical issues in our derivation.   The first is the choice of the vacuum. We have chosen as vacuum 
AdS$_3$ in Poincar\'e coordinates. This is the most natural choice in the context we are considering, because differently from  AdS$_3$ in global coordinates, the Poincar\'e vacuum is continuously connected both with the $T>0$ black hole spectrum and with the part of the spectrum describing conical singularities. Nevertheless, one could also consider the possibility of choosing AdS$_3$ in global coordinates as the vacuum. 

 A second, related, issue is represented by the Hawking-Page phase transition occurring when $r_+\sim L$. Below this point the BTZ black hole becomes unstable and AdS$_3$ at finite temperature becomes energetically preferred. 
Our cylinder approximations hold in the two limits $r_+>>L$ or $r_+<<L$, i.e far away from the critical point. Whereas the regime  $r_+>>L$ is well understood - the path integral for AdS$_3$ Euclidean gravity is dominated by the contribution coming from the BTZ black hole - what happens in the regime  $r_+<<L$  in presence of localized matter is not completely clear.
This point is also strongly related to the last issue, namely the fact that in principle singular geometries, like 
conical singularities, cannot be considered as allowed contributions to the Euclidean quantum gravity partition function. On the other hand,    this regime is expected to be outside the  domain of validity of the geometric spacetime description, i.e fully inside a would-be pre-geometric phase.

Our results  rely heavily on the peculiarity of 3D AdS gravity, we therefore expect generalization to four or higher dimensions to be rather involved. For instance, in 4D gravity we have instead of conical singularities generated by a pointlike positive mass, naked curvature singularities generated by localized sources with negative mass.

\end{document}